\documentstyle[12pt]{article}

\def\hybrid{\topmargin 0pt      \oddsidemargin 0pt
          \headheight 0pt \headsep 0pt
          \voffset=-0.5cm
          \textwidth 6.25in       
          \textheight 9.5in       
          \marginparwidth 0.0in
          \parskip 5pt plus 1pt   \jot = 1.5ex}
\catcode`\@=11
\def\marginnote#1{}

\newcount\hour
\newcount\minute
\newtoks\amorpm
\hour=\time\divide\hour by60
\minute=\time{\multiply\hour by60 \global\advance\minute by-\hour}
\edef\standardtime{{\ifnum\hour<12 \global\amorpm={am}%
          \else\global\amorpm={pm}\advance\hour by-12 \fi
          \ifnum\hour=0 \hour=12 \fi
          \number\hour:\ifnum\minute<10 0\fi\number\minute\the\amorpm}}
\edef\militarytime{\number\hour:\ifnum\minute<10 0\fi\number\minute}

\def\draftlabel#1{{\@bsphack\if@filesw {\let\thepage\relax
     \xdef\@gtempa{\write\@auxout{\string
        \newlabel{#1}{{\@currentlabel}{\thepage}}}}}\@gtempa
     \if@nobreak \ifvmode\nobreak\fi\fi\fi\@esphack}
          \gdef\@eqnlabel{#1}}
\def\@eqnlabel{}
\def\@vacuum{}
\def\draftmarginnote#1{\marginpar{\raggedright\scriptsize\tt#1}}

\def\draftlabel#1{{\@bsphack\if@filesw {\let\thepage\relax
     \xdef\@gtempa{\write\@auxout{\string
        \newlabel{#1}{{\@currentlabel}{\thepage}}}}}\@gtempa
     \if@nobreak \ifvmode\nobreak\fi\fi\fi\@esphack}
          \gdef\@eqnlabel{#1}}
\def\@eqnlabel{}
\def\@vacuum{}
\def\draftmarginnote#1{\marginpar{\raggedright\scriptsize\tt#1}}

\def\draft{\oddsidemargin -.5truein
          \def\@oddfoot{\sl preliminary draft \hfil
          \rm\thepage\hfil\sl\today\quad\militarytime}
          \let\@evenfoot\@oddfoot \overfullrule 3pt
          \let\label=\draftlabel
          \let\marginnote=\draftmarginnote
     \def\@eqnnum{(\theequation)\rlap{\kern\marginparsep\tt\@eqnlabel}%
\global\let\@eqnlabel\@vacuum}  }

\def\numberbysection{\@addtoreset{equation}{section}
          \def\theequation{\thesection.\arabic{equation}}}

\def\underline#1{\relax\ifmmode\@@underline#1\else
          $\@@underline{\hbox{#1}}$\relax\fi}

\def\titlepage{\@restonecolfalse\if@twocolumn\@restonecoltrue\onecolumn
       \else \newpage \fi \thispagestyle{empty}\c@page\z@
          \def\thefootnote{\fnsymbol{footnote}} }

\def\endtitlepage{\if@restonecol\twocolumn \else  \fi
          \def\thefootnote{\arabic{footnote}}
          \setcounter{footnote}{0}}  
\relax

\hybrid

\def\beq{\begin{equation}}
\def\eeq{\end{equation}}
\def\p{\partial}

\begin{document}
\begin{titlepage}

\title{$\tau$-function for analytic curves}

\author
{I.K.Kostov\thanks{Service de 
Physique Th\'eorique, CEA-Saclay,
91191 Gif sur Yvette, France}, 
\,
I.Krichever\thanks{Department of 
Mathematics, Columbia University, New York, NY 10027, USA and
Landau Institute for Theoretical Physics}, \,
M.Mineev-Weinstein\thanks{Theoretical Division, 
MS-B213, LANL, Los-Alamos,
NM 87545, USA},\,\\
P.B.Wiegmann\thanks{James 
Franck Institute and Enrico Fermi
Institute
of the University of 
Chicago, 5640 S.Ellis Avenue, Chicago, IL 60637, USA
and Landau Institute for Theoretical Physics},\,
A.Zabrodin\thanks{Joint 
Institute of Chemical Physics, Kosygina str. 4, 117334,
Moscow, Russia and ITEP, 117259, Moscow, Russia}}

\date{May 2000}
\maketitle

\begin{abstract}
We review the concept of 
$\tau$-function for simple  analytic curves. The
$\tau$-function gives 
a formal solution to the 2D inverse
potential problem and appears as 
the $\tau$-function of the integrable
hierarchy which describes 
conformal maps of simply-connected
domains bounded by analytic curves 
to the unit disk. The $\tau$-function
also emerges in the context of 
topological gravity and enjoys an
interpretation as a large $N$ limit 
of the normal matrix 
model.
\end{abstract}

\vfill

\end{titlepage}
\noindent
1. 
Recently, it has been realized
\cite{mwz,wz} that  conformal maps
exhibit an integrable  structure: conformal maps of 
compact simply connected 
domains bounded by 
analytic curves provide a solution to the 
dispersionless limit of
the 2D Toda hierarchy. As is well known 
from  the theory of solitons,
solutions of an integrable hierarchy 
are represented by $\tau$-functions.
The dispersionless limit of 
the $\tau$-function emerges as a natural object
associated with the curves.
In this paper we discuss the $\tau$-function for 
simple analytic curves and its connection to the 
inverse potential problem, 
area preserving  diffiomorphisms,  
the Dirichlet boundary problem, and matrix models.

\bigskip
\noindent
2. {\it Inverse potential problem.} 
Consider a closed analytic
curve\footnote{A closed
analytic curve is 
the curve which can be parametrized by
a function
$z\equiv 
x+iy=z(w)$, analytic in a domain which includes  the
unit circle 
$|w|=1$} $\gamma$ in the complex plane
and denote by $D_+$ and  $D_-$ 
the
interior and exterior
   domains with respect to the curve. The 
point $z=0$ is
assumed to be in $D_{+}$.
    Assume that the domain 
$D_+$ is filled homogeneously
    with  electric charge,  with a
 density which we set to be  equal to 1.
The potential $\Phi$ created 
by the charge obeys the
equation
\beq
\label{1}   - \p_{z}\p_{\bar z}\Phi(z,\bar z)   = \cases{{1 } &
   \quad if  \quad
   $z=x+iy \in 
D_+$  \cr
      0&  \quad if  \quad  $z=x+iy \in
D_-$\   \cr}
 \eeq
The  potential $\Phi$ can be  written as an integral over the 
domain $D^+$:
\beq
\label{1bis}
\Phi (z,\bar z)=-\frac{2}{\pi}\int_{D_{+}}d^2 z'\ \log|z-z'| 
\eeq
In the 
exterior domain $D_{-}$, the potential  is   the
   harmonic function 
whose asymptotic expansion as $z\to \infty$
is given 
by
\beq\label{2}\Phi^{-}(z,\bar z) =-2t_{0}\log|z|+
2{\cal 
R}e\sum_{k>0}\frac{v_k}{k} 
z^{-k},
\eeq
where
\beq\label{3}
v_k=\frac{1}{\pi}\int_{D_+} z^k d^2 z
    \qquad (k > 0)
\eeq
are  the harmonic moments of 
the interior domain $D_{+}$ and
\beq\label{3bis}
\pi t_0 
=\int_{D_{+}}d^2z
\eeq
   is its area.
   In the interior domain 
$D_+$, the  potential (\ref{1bis})
     is equal   to  a  function 
$\Phi^{+}$, which is harmonic up to
the term
   $ -|z|^2$. The 
expansion of this function around $z=0$ 
is
\beq\label{4}
\Phi^{+}(z,\bar z)= -|z|^2- v_0+
2{\cal 
R}e\sum_{k>0}t_k z^k
\eeq
Here
\beq\label{5}
t_k=-\, \frac{1}{\pi 
k}\int_{D_-}z^{-k}d^2 z \qquad (k > 0)
\eeq
are the harmonic moments 
of the exterior domain $D_{-}$ and
\beq 
\label{v0}
v_0=\frac{2}{\pi}\int_{D_+}  \log|z|\,d^2z.
\eeq
The two sets of moments (\ref{3}) 
and
(\ref{5}) are related by the  conditions that 
$\Phi^+=\Phi^-$,
$\p_{z} \Phi^+ =\p_{z} \Phi^-$
on the curve $\gamma$.

The inverse potential problem is to determine 
the form of the   curve
$\gamma$  given one of the functions $\Phi^+$ 
or $\Phi^-$,
i.e. given one of the infinite sets of moments.
   We 
will choose as independent
variables the
area $\pi t_0$ and the 
moments of
the exterior  $t_k \ (k\ge 1)$. Under certain 
conditions, they completely determine  
the form of
the curve as well as the moments $v_k \ (k\ge 0)$ 
\cite{Brodsky}.
More precisely, $\{t_k\}_{k=0}^{\infty}$  is a good set 
of local coordinates
in the space of analytic curves. For simplicity 
we assume
in this paper that only a finite number of $t_k$ are 
non-zero.
In this case the series (\ref{4}) is a polynomial in $z$, 
$\bar z$
and, therefore, it gives the function $\Phi^{+}$ for $z\in 
D_{+}$.
Note that $t_0$, $v_0$ are real
quantities while all other 
moments are in general complex variables.

\bigskip
\noindent
3. {\it 
Variational principle}.
Consider the energy functional
describing a 
charge with a density $\rho (z,\bar z)$ in the background
potential 
created by the homogeneously distributed charge with the
density +1 
inside the domain $D_+$ (\ref{1}):
\beq\label{32B}
{\cal 
E}\{\rho\}=-\frac{1}{\pi^2}
\int\! \int d^2 z \,d^2 z' \  \rho(z, 
\bar z)\  \log|z-z'|\rho(z',\bar
z')\
\,
   -\frac{1}{\pi}\int\! d^2 
z\ \rho(z,\bar z) \ \Phi(z,\bar z).
\eeq

The first term is the 2D 
``Coulomb'' energy of the charge
while the second one is the energy 
due to the background charge.
Clearly, the distribution of the 
charge
   neutralizing the background charge
gives the minimum to 
the
functional: $\rho_{0}=-1$ inside the domain and $\rho_{0}=0$ 
outside. At
the minimum the functional is equal to minus 
electrostatic energy
$E$ of the
background charge :
\beq
\label{32A}
-E={\rm \min}_{\rho} {\cal E}\{\rho\}  =
\frac{1}{\pi^2}
\int_{D_{+}} \!\!\!\! d^2 z  \int_{D_{+}}\!\!\!\!d^2 
z' \ \log|z-z'|
=-\frac{1}{2\pi} \int_{D_{+}} \!\!\!\! d^2 z \ 
\Phi(z,\bar z).
   \eeq
Varying over $\rho$ and then setting 
$\rho=-1$ inside the domain, we
obtain eq.(\ref{4}).

The first 
corollary of the variational principle
is that the $E$
is a potential 
function for the moments.
Eq.\,(\ref{4}) suggests to treat $v_0$ and 
$t_k$
as independent variables, so
moments of the interior, $v_k$, 
$k\geq 1$, and $t_0$ are
functions of $v_0$ and
$t_k$. Let us 
differentiate $E$ or
$-{\cal E}\{\rho \}$ at the extremum with 
respect to the parameters
$v_0$, $t_k$. Since
$\rho_{0}$ minimizes 
the functional, the derivative is equivalent to the
partial 
derivative of
${\cal E}$ at the fixed extremum
$\rho$. This 
gives
\beq\label{25a}
\frac{\p E}{\p t_k}= v_k, \;\;\;
\frac{\p E}{\p 
\bar t_k}= \bar v_k,
\;\;\;
\frac{\p E}{\p v_0}=- t_0,
\eeq
where the 
partial derivative with respect to
$t_k$ is taken at fixed $v_0$ and 
$t_j \ (j   \neq 0, k)$.
Therefore the differential $dE$ 
reads
\beq\label{dE}
dE= \sum_{k>0}(v_k dt_k +\bar v_k d\bar t_k )
- t_0 dv_0.
\eeq

Let us note that the variational principle may be 
formulated in a
number of different ways. One particular 
variational
principle is suggested by the matrix model discussed in 
the Sec.9. In
this case one consider a charged liquid in the 
potential
\beq
\label{bcgr}
V(z, \bar z)=  z \bar z +v_0-
 \sum_{k>0}\left ( t_k z^k + \bar t_k \bar z^k \right )
\eeq
defined 
everywhere on the plane and $v_0$ and $t_k$ are parameters.
The 
energy of the charged liquid
\beq\label{32}
{\cal E}\{\rho, 
V\}=-\frac{1}{\pi^2}
\int\!d^2 z \! \int d^2 z' \  \rho(z, \bar z)\ 
\rho(z',\bar z')\ \log|z-z'|
\,+\frac{1}{\pi}\int\! d^2 z\ 
\rho(z,\bar z) \ V(z,\bar z).
\eeq
reaches the minimum  if the liquid 
forms a drop with the density
$\rho_0=-1$ bounded by the curve 
determined by parameters of the
potential $v_0$ and $t_k$. For 
another version of the variational
principle see 
\cite{Ivanov}.

\bigskip
\noindent
4.    {\it $\tau$-function.}
\,It 
is more natural to treat the total charge $t_0$
rather than $v_0$ as 
an independent variable, i.e. to consider the
variational principle 
at a fixed total charge $t_0=\int\rho d^2z$. This
is achieved via the 
Legendre transformation. Let us introduce the function
$F=E+ t_0 v_0$ 
,  whose  differential is
\beq\label{dF}
dF= \sum_{k>0}(v_k dt_k 
+\bar v_k d\bar t_k )
+ v_0 dt_0.
\eeq
We define the $\tau$-function 
as $ \tau =e^F$, so that
\beq\label{31}
\ \log\tau
=\frac{1}{2\pi} 
\int_{D_{+}} \!\!\!\! d^2 z \ \Phi(z,\bar z)  + t_0v_0\ =\ 
-\frac{1}{\pi }
\int \!\int_{D_{+}} \log \left |
\frac{1}{z}-\frac{1}{z'}
\right |d^2 z d^2 z'.
\eeq

The 
$\tau$-function is a real function
of the moments
$\{t_0 , \, t_1 , 
\, t_2 , \ldots \}$.
Under the assumption that only a finite number 
of them
are non-zero, we can substitute (\ref{4}) into (\ref{31})
and 
perform the term-wise integration.
Taking into account 
that
$
\frac{1}{\pi}\int_{D_{+}} |z|^2 d^2z =
\frac{1}{2}t_0^2 
+\frac{1}{2}
\sum _{k>0}k(t_k v_k +\bar t_k \bar v_k)$
(a simple 
consequence of the Stokes formula),
we get the
expression for 
the
$\tau$-function in terms of $t_k$ and 
$v_k$:
\beq
\label{free2}
2\log \tau = -\frac{1}{2} t_0^2 +t_0 v_0 -
\frac{1}{2}\sum_{k>0}(k-2)(t_k v_k +\bar t_k \bar v_k).
\eeq

Rephraising (\ref{25a}) we get the main property of the
$\tau$-function, which has been used as its definition in Ref.\cite{wz}
\beq\label{25}
\frac{\p \log\tau}{\p t_k}=v_k, \;\;\:
\frac{\p \log\tau}{\p \bar t_k}=\bar v_k,
\;\;\;
\frac{\p \log\tau}{\p t_0}=v_0
\eeq
where the derivative with respect to $t_k$ is taken at
fixed $t_j$ ($j \neq k$).

Two immediate consequences of the very
existence of the potential function  are
symmetry relations for the moments
\beq\label{15}
\frac{\p v_k}{\p t_n}=\frac{\p v_n}{\p t_k},
\;\;\;\;\;\;
\frac{\p v_k}{\p \bar t_n}=\frac{\p \bar v_n}{\p t_k}
\eeq
and the quasi-homogeneity condition for the $\tau$-function:
\beq
4\log\tau=
- t_0^2+ 2 t_0\frac{\p \log\tau}{\p t_0}
- \sum_{n>0}(n-2) \Bigl (t_n\frac{\p \log\tau}{\p t_n}+
\bar t_n \frac{\p \log\tau}{\p \bar t_n}\Bigr ).
\eeq
Apart from the term 
$- t_0^2$, this formula reflects the scaling
of moments as $z\to \lambda z$: 
$t_k \to \lambda^{2-k}t_k$ ($k\geq 0$),
$v_k \to \lambda^{2+k}v_k$ ($k \geq 1$).

As an illustration we present the $\tau$-function of ellipse
\cite{wz}. In this case only the first two moments $t_1$ and $t_2$
are nonzero\footnote{The $\tau$-fuction for the ellipse (at $t_1=0$)
appeared in Ref.\,\cite{FGIL} as the limit 
of the Laughlin wave function or a planar
limit of the free energy of normal matrix models, see Sec.\,9}:
$$
\mbox{log}\tau= -
\frac{3}{4}t_{0}^2
+ \frac{1}{2}t_{0}^2\ \mbox{log}\, 
\left ( {t_{0}\over 1-4|t_2|^2)}\right )  +
{t_{0} \over 1-4|t_2|^2} \left (
|t_1|^2+t_1^2\bar t_2+\bar t_1^2 t_2 \right ).
$$

\bigskip\noindent
5.   {\it  Schwarz function and generating function of the conformal
map.}\  Consider a univalent conformal map of the exterior
domain $D_-$ to the exterior of the unit disk and expand it
in  Laurent series:
\beq
\label{rwz}
w(z)=\frac{1}{r}z+\sum_{j=0}^{\infty}p_j z^{-j},
\eeq
where the coefficient $r$ is chosen to be real and
positive. The series for the inverse map (from
the exterior of the unit disk
to $D_-$) has a similar form:
\beq
\label{z}
z(w)=r w+\sum_{j=0}^{\infty}u_j w^{-j}.
\eeq
Chosen $w$ on 
the unit circle, eq.(\ref{z}) gives a
parametrization of the curve. 
By the definition of an
analytic curve, the map can be analytically 
continued
to a strip-like neighborhood of the curve belonging to 
$D_{+}$.
The continuation is given by the Riemann-Schwarz
reflection 
principle (see e.g.\cite{GC}):
\beq\label{41}
w=(\bar 
w(S(z)))^{-1},
\eeq
where $S(z)$ is the point reflected relative to 
the
curve\footnote{We use  the notation:
given an analytic 
function
$f(z)=\sum_j f_j z^j$, we set
$\bar f(z)=\sum_j \bar f_j 
z^j$.}.
Following \cite{Davies}, we call $S(z)$ {\it the Schwarz 
function}
of the curve.
Let us recall its construction. Write the 
equation for
the curve $F(x,y)=0$ in complex 
coordinates,
$F(\frac{z+\bar z}{2},\frac{z-\bar z}{2i})=0$, and solve 
it
with respect to $\bar z$. One gets the Schwarz function:
$\bar 
z=S(z)$. The Schwarz function is analytic in a
strip-like domain 
that
includes the curve. On the curve the Schwarz function is
equal 
to the complex conjugate argument. The main property
of the Schwarz 
function is the obvious but
   important {\it unitarity 
condition}
\beq
\label{c2}
\bar S(S(z))=z
\eeq
(the inverse function 
coincides with the complex conjugate
function).
In terms of a 
conformal map the Schwarz function is
\beq\label{55}
S(z)=r 
w^{-1}(z)+\sum_{j=0}^{\infty}\bar u_j w^j(z).
\eeq
Using the Schwarz 
function one can write the moments of the
exterior and the interior 
domains (\ref{3},\ref{5}) as
contour integrals \footnote {This is due 
to a more general statement $
\int_{D_\pm}f(z)d^2z=\pm\frac{1}{ 2i}
\oint_{\gamma} f(z)S(z)dz ,$
where $f(z)$ is an analytic function 
in the domain $D_{\pm}$.}
\beq\label{42}
t_n=\frac{1}{ 2\pi in}\oint_{\gamma} z^{-n}S(z) d z ,\;\;\;\;
v_n=\frac{1}{2\pi i}\oint_{\gamma} z^{n}S(z)d z 
\eeq
Eq.  (\ref{42}) yields the 
Laurent expansion of the
Schwarz 
function
\beq
\label{lg7}
S(z) =\sum_{k=1}^\infty k 
t_{k}z^{k-1}+\frac{t_0}{z}+\sum_{k=1}^\infty v_k z^{-k-1}.
\eeq

Now 
let us define  the {\it generating function} $\Omega (z)$,
related to 
the Schwarz function  by
\beq
\label{genf}
S(z)=\p_z\Omega (z).
\eeq
 
The latter is given, according to
   (\ref{lg7}), by the
    Laurent 
series
\beq
\label{HJ4}
\Omega (z) \ =\ \sum_{k=1}^{\infty} t_k z^k 
-\frac{1}{2}v_0
+t_0\,\mbox{log}\,z-
\sum_{k=1}^{\infty}\frac{v_k}{k} z^{-k}
\eeq
It can be represented as
$\Omega 
(z)=\Omega^{(+)}(z)+
\Omega^{(-)}(z)-\frac{1}{2}v_0$, where 
$\Omega^{(\pm )}(z)$
are analytic in $D_{\pm}$ 
respectively:
\beq
\label{lg8}
\Omega^{(+)}(z)=\frac{1}{\pi}
\int_{D_- }
\mbox{log}\Bigl ( 1-\frac{z}{z'} \Bigr ) d^2 z'
=\sum_{k=1}^{\infty} t_k  z^k
\eeq
\beq
\label{lg81}
\Omega^{(-)}(z)= 
\frac{1}{\pi}
\int_{D_+}
\log ( z-z') d^2 z'=t_0\log z
-\sum_{k=1}^{\infty}\frac{v_k}{k}z^{-k}
\eeq
  From 
(\ref{2},\ref{4}) we see that
$\Phi^- (z,\bar z)= - 2{\cal R}e \, 
\Omega^{(-)}(z)$ and $\Phi^+ (z,\bar
z)=  2{\cal R}e \, 
\Omega^{(+)}(z)-v_0-|z|^2$. Contrary to the potentials
$\Phi^{\pm}$, 
the analytical functions $\Omega^{+}$ and $-\Omega^{-}$ do
not match 
each other on the curve. The discontinuity gives the value of
the 
generating function restricted to the curve
\beq
\label{reim}
\Omega 
(z) =\frac{1}{2}|z|^2+2 i A(z),\;\;\;\;\;z\in\gamma
\eeq
where $A(z)$ 
is the area of the
interior domain bound by the ray
$\varphi 
=\mbox{arg}\,z$ and
the real axis.
   As a corollary, it is easy to 
show
that  variations of the
$\Omega (z)$ on the curve with respect 
to the {\it real}
parameters $t_0$, ${\cal R}e\, t_k$ and ${\cal 
I}m\, t_k$ are purely
imaginary. This allows one to apply the 
Riemann-Schwarz
reflection principle to analytical continuation 
of
\beq\label{44}
H_k(z)=\p_{t_k}\Omega(z),\;\;\;\bar
H_k(z)=-\p_{\bar 
t_k}\Omega(z)
\eeq
and to prove the fundamental 
relations
\beq\label{45}
\p_{t_0}\Omega(z)=\log 
w(z),
\eeq
\beq
\label{46}
\p_{t_k}\Omega(z)=\Bigl (z^k(w)\Bigr )_{+} 
+\frac{1}{2}
\Bigl (z^k(w) \Bigr )_{0}
\eeq
\beq
\label{47}
\p_{\bar 
t_k}\Omega(z)=\Bigl (S^k(z(w))\Bigr )_{-}
+\frac{1}{2}
\Bigl 
(S^k(z(w)) \Bigr )_{0}
\eeq
The symbols $(f(w))_{\pm}$  mean a 
truncated Laurent series, where
only  terms with positive (negative)
powers of $w$ are kept, while $(f(w))_{0}$ is the
constant term ($w^0$) of the series. Note that the
derivatives in eqs.(\ref{45}-\ref{47}) are taken
at fixed $z$.

To prove (\ref{45}), we first notice that
$$\p_{t_0}\Omega(z(w))=\log
z-\frac{\p_{t_0}v_0}{2}+\mbox{negative powers in}\; z=\log
w r-\frac{\p_{t_0}v_0}{2}+\mbox{negative powers
in}\;w.$$ Independently, one can show that
$\p_{t_0} v_0=2\,\mbox{log} \,r$.

Then,  using
the Riemann-Schwarz reflection principle, we may write
$\p_{t_0}\Omega(z(w))$ also in the form
$\p_{t_{0}}\bar\Omega(S(z(w))$. Expanding the latter in $S(z)$ and
then, using expansion of (\ref{55}) in $w$, we have
$$\p_{t_0}\bar\Omega(S(z(w))=\log
S(z)-\frac{\p_{t_0}v_0}{2}+\mbox{negative powers in}
\;S(z)=\log w+\mbox{positive powers in}\; w.$$
Comparing both
expansions, we conclude that  $\p_{t_0}\Omega(z)=\log w(z)$.
Similar arguments are used in the proof of (\ref{46}) and
(\ref{47}).

\bigskip
\noindent
6. {\it  Dispersionless Hirota equation and the Dirichlet boundary
problem}.
\   Using the representation (\ref{25}) of the moments $v_{k}$ as
derivatives of the $\tau$-function, one can express
the conformal map $w(z)$ (\ref{45})
through the  $\tau$-function:
\beq
\label{i}
\mbox{log}\,w=\mbox{log}\,z-\p_{t_0}\left (
\frac{1}{2}\p_{t_0}+\sum_{k\geq 1}
\frac{z^{-k}}{k} \p_{t_k}\right )
\,\mbox{log}\,\tau.
\eeq
With the help of the $\tau$-function,
eqs.(\ref{46},\ref{47}) can be similarly encoded
as follows:
\beq
\label{log1}
\p_{z}\p_{\zeta}\,\mbox{log}\,\Bigl (
w(z)-w(\zeta )\Bigr ) =
\frac{1}{(z-\zeta )^2} +
\left ( \sum_{k\geq 1}z^{-k-1}\p_{t_k}\right )
\left ( \sum_{n\geq 1}\zeta ^{-n-1}\p_{t_n}\right )
\,\mbox{log}\,\tau
\eeq
\beq
\label{log2}
- \,\p_{z}\p_{\bar \zeta}\,\mbox{log}\,\Bigl (
w(z)\bar w(\bar \zeta ) -1 \Bigr ) =
\left ( \sum_{k\geq 1}z^{-k-1}\p_{t_k}\right )
\left ( \sum_{n\geq 1}\bar \zeta ^{-n-1}\p_{\bar t_n}\right )
\,\mbox{log}\,\tau
\eeq
The derivation is similar to the
one given 
in \cite{GiKo1,KC} for the case of the KP
hierarchy.  
Moreover, 
these equations in the integrated form 
are most conveniently written 
in terms of the 
differential
operators
\beq
\label{DDD}
D(z)=\sum_{k\geq 
1}\frac{z^{-k}}{k}\p_{t_k}\,,
\;\;\;\;\;\;
\bar D(\bar z)=\sum_{k\geq 
1}\frac{\bar z^{-k}}{k}\p_{\bar t_k}
\eeq
From 
(\ref{log1},\ref{log2}) one obtains:
\beq
\label{51}
\mbox{log}\, 
\frac{w(z)-w(\zeta )}{z-\zeta}\, =\,
- \, 
\frac{1}{2}\p_{t_0}^2\,\mbox{log}\,\tau+
D(z)D(\zeta  )
\,\mbox{log}\,\tau
\eeq
\beq
\label{511}
-\, \mbox{log}\, \left (1- 
\frac{1}{w(z)
\bar w (\bar \zeta )}\right )=
D(z)\bar D(\bar \zeta )
\,\mbox{log}\,\tau
\eeq

Combining  (\ref{i}) and (\ref{51}), one 
obtains the
dispersionless Hirota equation
    (or the dispersionless  Fay identity) for 2D Toda lattice hierarchy
      \cite{wz}:
\begin{equation}
\label{Hirota}
(z\! -\! \zeta)e^{ D(z)D(\zeta )
\, \log \tau } =
\! z e^{
- \p_{t_0}D(z) \log \tau } -
\!\zeta e^{
- \p_{t_0}D(\zeta ) \log \tau }
\end{equation}
Eq.\,(\ref{Hirota}), after being expanded in
powers of $z$ and $\zeta$,
generates an infinite set of relations
between the second derivatives
$\p_{t_n}\p_{t_m}\mbox{log}\tau$ of the $\tau$-function.
Using (\ref{511}) instead of (\ref{51}), a similar
equation for the mixed derivatives
$\p_{t_n}\p_{\bar t_m}\mbox{log}\tau$ can be written:
\beq
\label{Hirota1}
1- e^{ -D(z)\bar D(\bar \zeta ) \log \tau } =
\frac{1}{z\bar \zeta } \, e^{
\p_{t_0} (\p_{t_0} +D(z)+\bar D(\bar \zeta )) \log \tau }
\eeq
 
Let us conclude this section with two other forms
of the dispersionless Hirota equation
for the conformal map. They emphasize a relation between
the Hirota equation and two  fundamental objects of the classical
analysis: the Green function of the Dirichlet problem\footnote{This
relation is pointed out to us by L. Takhtajan.} and the Schwarz
derivative.

The Green function of the Dirichlet boundary problem
for the Laplace operator in $D_{-}$ expressed through
the conformal map $w(z)$ is:
\beq
\label{Green}
G(z,\zeta )=\,\mbox{log}\,\left |
\frac{w(z)-w(\zeta )}{w(z)\bar w(\bar \zeta )-1}\right |
\eeq
Combining (\ref{51}) and (\ref{511}), and using the
notation (\ref{DDD}), we represent the Green function
as follows:
\beq
\label{Green1}
2G(z,\zeta )=2\, \mbox{log}\, |z^{-1}-\zeta ^{-1}|
+\Bigl (\p_{t_0}+D(z)+\bar D(\bar z)\Bigr )
\Bigl (\p_{t_0}+D(\zeta )+\bar D(\bar \zeta )\Bigr )
\,\mbox{log}\,\tau
\eeq
This formula generalizes (\ref{i}) since (\ref{Green1})
becomes the real part of (\ref{i}) as $\zeta \to \infty$.
(As $\zeta \to \infty$, $G(z,\zeta )\to -\,\mbox{log}\, |w(z)|$.)
Note also that the real part of (\ref{i}) can be written
in the form
$$
\Phi (z, \bar z)= -2 t_0 \, \mbox{log}\, |z|
+ \Bigl ( D(z) + \bar D (\bar z) \Bigr )
\,\mbox{log}\,\tau
$$
where 
$\Phi$ is the potential (\ref{1bis}) ($z\in D_-$). 

The l.h.s. of 
eq.(\ref{51}) generalizes the Schwarz derivative of
the conformal 
map
\beq
\label{sd}
T(z)\equiv\frac{w'''(z)}{w'(z)}-\frac{3}{2}\left 
(
\frac{w''(z)}{w'(z)}\right )^2=6 \ 
\mbox{lim}_{_{ 
z\to\zeta}}\p_z\p_\zeta
\mbox{log}\, \frac{w(z)-w(\zeta )}{z-\zeta }
\eeq
    Taking the limit $\zeta\to z$ of both sides of 
(\ref{51}), we get a
relation between the Schwarz derivative and the 
$\tau$-function:
\beq\label{Schwder}
T(z)=6\,z^{-2}\sum_{k,n\geq 
1}z^{-k-n}
\frac{\p^2 \mbox{log}\,\tau}{\p t_k \p t_n}
\eeq
The latter 
can be used as an alternative definition of the 
$\tau$-function.

\bigskip
\noindent
7.    {\it Integrable structure 
of conformal maps.}\ 
Eqs.(\ref{45}-\ref{47}) 
allow one to say that the differential
\begin{equation}
\label{95}
d\Omega 
=Sdz+\,\mbox{log}\,
     w \ dt_{0} +\sum_{k=1}^\infty(H_k dt_k-\bar 
H_k d\bar t_k)
\end{equation}
generates the set 
of Hamiltonian equations for deformations of
the curve due to variation of $t_k$:
\beq\label{61}
\p_{t_k}S(z)=\p_z H_k(z),\;\;\;\;\p_{\bar 
t_k}S(z)=-\p_z
\bar H_k(z),
\eeq
where we set $H_0(z)=\log w(z)$. The 
equations are consistent
due to commutativity of the 
flows:
\begin{equation}
\label{62}
\left ( \p_{t_j}  H_k\right )_z
=\left ( \p_{t_k}  H_j\right )_z=\p_{t_j}\p_{t_k}\Omega(z)
\end{equation}
Equations (\ref{61}) 
are more transparent being written
in terms of canonical variables. 
The differential $d\Omega$
suggests that the  pairs $\log w,\; t_{0} 
$ and $z(w),\;S(z(w))$ are
canonical and establishes {\it the 
symplectic structure for
conformal maps}. Indeed, treating
$w$ as an 
independent variable, one rewrites eq.\,(\ref{45}) as
\begin{equation}\label{pb}
\{z(w),\,S( 
z(w))\}=1
\end{equation}
where  the Poisson bracket $\{,\} $ is with 
respect to $\log
w$ and the area $t_0$ is defined 
as
\beq
\label{PB}
\{f,\,g\}=w\frac{\p f}{\p w}\frac{\p g}{\p t_{0}}
-w\frac{\p g}{\p w}\frac{\p f}{\p t_{0} }
\eeq
where the derivatives 
with respect to $t_0$ are taken at fixed  $t_k$ and $w$.

The other flows read
\beq
\label{63}
\frac{\p z(w)}{\p t_k} =\{H_k, 
z(w)\}
\eeq
\beq
\label{64}
\frac{\p S(z(w))}{\p t_k} =\{H_k, 
S(z(w)\},
\eeq
and similarly for the flows with respect to $\bar 
t_k$.
Now the Hamiltonian functions $H_k$ and $\bar H_k$ are 
degree
$k$ polynomials of $w$ and $w^{-1}$ respectively.

The consistency conditions (\ref{61}) now take the form
of the zero-curvature conditions:
\beq\label{65}\p _{t_j} H_i
-\p _{t_i}H_j 
+\{H_i,\,H_j\}=0,
\eeq
\beq\label{66}\p _{t_j}\bar H_i
+\p _{\bar 
t_i}H_j +\{\bar H_i,\,H_j\}=0.
\eeq
The infinite set of the Poisson-commutating flows 
form a {\it Whitham integrable hierarchy} \cite{kr2}. 
Eqs.\,(\ref{63},\ref{64})
are the Lax-Sato equations for the 
hierarchy. They generate an infinite
set of differential equations for the 
coefficients (potentials) $u_j$ of the inverse conformal map 
(\ref{z}).
The first equation of the hierarchy is
\beq\label{67}
\p_{t_1\bar 
t_1}^2\phi=\p_{t_0}\exp(\p_{t_0}\phi),\;\;\;\;\;
\p_{t_0}\phi=\log r^2.
\eeq
The integrable hierarchy describing conformal maps 
is also known in the soliton literature 
as the {\it  dispersionless
Toda lattice hierarchy}, or SDiff(2) Toda hierarchy 
\cite{Ta1} (see the next 
section)\footnote{
A relation between conformal maps of slit 
domains and special solutions to equations of 
hydrodynamic type 
(Benney equations) was first observed
by Gibbons and Tsarev \cite{GT1} }.
The algebra Sdiff(2) of area-preserving diffeomorphisms
is the symmetry algebra of this hierarchy \cite{Ta1}.
Eqs.\,(\ref{63}-\ref{66}) 
describe infinitesimal deformations of the curve
such that the area $t_0$ is kept fixed.

The  integrable hierarchy possesses many 
solutions. The
particular solution relevant to conformal maps is 
selected by  the
subsidiary  condition  (\ref{pb}). This condition, 
known as   {\it
dispersionless string equation},  has already 
appeared in the study of the
$c=1$ topological gravity 
\cite{Ta1,gravity1,AoKo}  and
in the large $N$ limit of a model of 
normal random matrices \cite{nmatrix}. 
The latter is discussed in Sec.\,9.

\bigskip
\noindent
8. {\it Toda lattice hierarchy and its 
dispersionless
limit}. \ Below we review the two dimensional Toda 
lattice
hierarchy and show that its dispersionless limit gives the
equations describing conformal maps
(\ref{46},\ref{47},\ref{63},\ref{64}).

The 2D Toda hierarchy is defined by two Lax operators
\beq
\label{L}
L=r(t_0)\ e^{\hbar\frac{\p}{\p{t_0}}}+
\sum_{k=0}^{\infty}u_k(t_0)\ e^{-k\hbar\frac{\p}{\p t_0}}
\eeq
\beq
\label{L1}
\bar L=e^{-\hbar\frac{\p}{\p{t_0}}}\ r(t_0 )+
\sum_{k=0}^{\infty}
e^{k\hbar\frac{\p}{\p{t_0}}}\bar u_k(t_0)
\eeq
acting in the space of functions of $t_0$
where the coefficients $u_j$ and $\bar u_j$ are functions of
$t_0$ and also of two independent sets of parameters
(``times")
$t_k$ and $\bar t_k$. Note that $u_k$ and $\bar u_k$ as
well as $t_k$ and $\bar t_k$ in (\ref{L},\ref{L1}) 
are not necessarily complex
conjugate to each other, although we  choose them to be so.

The dependence of the coefficient $u_k$ and $\bar u_k$ on
$t_k$ and $\bar t_k$ are given by the Lax-Sato equations:
\beq
\label{73}
\hbar\frac{\p L}{\p t_k} =[H_k, L]
\eeq
\beq
\label{74}
\hbar\frac{\p L}{\p \bar t_k} =[L,\bar H_k]
\eeq
and similar equations for $\bar L$.
The flows are generated by
\beq\label{75}
H_k =\bigl (L^k \bigr )_{+}+\frac{1}{2}\bigl (L^k \bigr )_{0}
\eeq
\beq\label{175}
\bar H_k =\bigl (\bar L^k \bigr )_{-}+\frac{1}{2}\bigl
(\bar L^k
\bigr )_{0}
\eeq
where the symbol $ \bigl (L^k \bigr )_{\pm}$ means positive
(negative) parts of the series in the shift operator
$e^{\hbar\frac{\p}{\p t_0}}$.
The first equation of the hierarchy is the Toda lattice equation
\beq\label{173}
\p_{t_1\bar
t_1}^2\phi(t_0)=e^{\phi(t_0+\hbar)-\phi(t_0)}-e^{\phi(t_0)-\phi(t_0-\hbar)},
\eeq
where $r^2=e^{\phi(t_0+\hbar)-\phi(t_0)}$.

The spectrum of the
Lax operator is determined by the linear problem
$L\Psi=z\Psi$. The wave function $\Psi$
is expressed through the $\tau$-function $\tau_{\hbar}$
of the dispersionfull hierarchy (\ref{73},\,\ref{74}) by
the following formula:
\beq\label{tau}
\Psi(z;t_0,t_1,t_2,\cdots)=\tau^{-1}_{\hbar}(t_0,t_1,t_2,\ldots)
z^{t_0 /\hbar} e^{\frac{1}{\hbar}\sum_{k>0}t_k z^k}
e^{\hbar \sum_{k>0}\frac{z^{-k}}{k}
\frac{\p}{\p{t_k}}}
\tau_{\hbar}(t_0,t_1,t_2,\ldots)
\eeq

Among many solutions of the hierarchy, one is of particular interest.
It is  selected
by the {\it string equation} \cite{Douglas}
\beq\label{string}[L,\bar L]=\hbar
\eeq
This solution is known to describe the normal matrix 
model at finite size
of matrices \cite{nmatrix}.

The dispersionless limit of the Toda hierarchy is a
formal semi-classical limit
$\hbar\to 0$. To proceed we notice that the shift operator
$W=e^{\hbar\frac{\p}{\p t_0}}$ obeys the commutation relation
$[W,\,t_0]=\hbar W$.
In the semiclassical limit it is supposed to be replaced
by the canonical variable $w$
with the Poisson bracket
$\{\mbox{log}w,t_0 \}=1$. The Lax operator then becomes
a $c$-valued function
which is identified with the inverse conformal map $z(w)$ (\ref{z}).
Similarly,  $\bar L$ is identified with $S(z(w))$.
In their turn, the Lax-Sato
equations (\ref{73},\ref{74}) are identified
with eqs.(\ref{63},\ref{64}) for
the conformal map. In the same fashion the dispersionless limit of the
string equation (\ref{string}) is identified  with  eq.(\ref{pb}). The
semiclassical limits of the wave function
and the $\tau$-function
give the generating function $\Omega$
and the dispersionless $\tau$-function:
$\Psi\to e^{\Omega/\hbar}$,
$\tau_{\hbar} \to e^{(\log\tau )/\hbar^{2}}$.
Similarly, eq. (\ref{Hirota}) is a  semiclassical limit of the Hirota
equation for the $\tau$-function of the 2D Toda hierarchy.

\bigskip\noindent
9.  {\it The $\tau$-function  of the conformal map
as large $N$ matrix integral.}
        The integrable structure of conformal maps is identical to the one
    observed in a class of random matrix models related to noncritical
    string theories. Moreover, there exists
      a  random matrix model  whose large $N$ limit
    reproduces {\it exactly} the $\tau$-function for 
 analytic curves.

    Consider  the partition function of the
    ensemble of normal
    random $N\times N$ 
matrices \cite{nmatrix}\footnote{Earlier V. Kazakov
  pointed to us  that the Lax equations (\ref{73},\ref{74}) are
generated by the Hermitian 2-matrix 
model \cite{2matrix} with complex
conjugated potentials. The latter and the normal 
matrix model have an
identical $1/N$-expansion.}, 
with the potential (\ref{bcgr}):
     \beq\label{tauM}
    \tau_{\hbar} [t , \bar t ]
     = \int
     dM d M^{\dag}
e^{ -{1\over \hbar} {\rm Tr}V(M, M^{\dag})}
\eeq
A  matrix is called  normal
    if it commutes with its Hermitian conjugated $[M , M^{\dag}]=0$.
Passing to the eigenvalues
    ${\rm diag}(z_{1}, \ldots, z_{N} )$ of the 
matrix $M$, one obtains the
    measure of the integral in a
    factorized form
    $ dM d M^{\dag} \sim  \  \prod _{i=1}^{N}
dz_{i}
    d\bar z_{i} \ \prod _{k<j} (z_{k } - z_{j } )(\bar z_{k } -
    \bar z_{j } )$. Then  the  partition function
is represents a two-dimensional Coulomb gas in the
potential (\ref{bcgr})
    \beq\label{COUL}
    \tau_{\hbar}  [t , \bar t ] =
    \int \prod_{k=1}^{N} dz_{k} d \bar z_{k}\
    e^{  - {1\over \hbar}
    V(z_{k}, \bar z_{k})}\
     \ \prod _{i<j} e^{2\log|z_{i } - z_{j } |}.
\eeq
    To proceed to the large $N$ limit one introduces a
parameter $t_{0} = \hbar N$ 
and expresses the integrand in terms of density of 
eigenvalues
as $e^{-\hbar^{-2}{\cal E}\{\rho, V\}}$,
where ${\cal E}\{\rho, V\}$
is given by eq. (\ref{32}). Then, the large $N$ 
($\hbar \to 0)$  limit
yields to the variational principle of Sec.3. 
In the  large $N$ limit
the eigenvalues of the matrix homogeneously 
fill the
domain $D_+$ bound
by the curve, characterized by the 
harmonic moments $t_k$ and the area
$t_0$ and leads to the 
$\tau$-function defined by eq.\,(\ref{31}).
Other objects introduced 
in Secs.\,3-7 can also be identified with
expectation values of the 
matrix model. In particular the moments $v_k$
(eq.(\ref{3})) 
are
$$v_k=\hbar\left\langle {\rm Tr}M^k\right\rangle$$
and 
$\Omega^--\frac{1}{2}v_0=\hbar
      \left\langle {\rm Tr} \log (z-M) 
\right\rangle $.

In order to identify the Lax operator, we 
follow \cite{Mehta,nmatrix,2matrix}.
Introduce   the basis of 
orthogonal
polynomials $ P_n(z) = h_{n}z^{n}+\ldots (n\ge 0),$  by 
the
orthonormality relations
\beq\label{ortP}\langle m |n \rangle 
\equiv  \int {d^2z   }
   \ \overline{ P_{n} (z)} \ e^{-{1\over \hbar} V(z,\bar z)}
\ P_{m}(z)=  \delta_{m,n}
\eeq
   The polynomials are uniquely defined by the potential $V$
 up to phase factors.
   It is easy to see that the   $\tau$-function 
is given
   by the product of the coefficients $N! h_{n}h_{n-1}\ldots 
h_{0}$ of the
highest powers of the polynomials 
$P_n(z)=h_nz^n+\cdots$.
   Then  Lax operators $L$ and $\bar L$ 
appear as  the operators $\langle
m|z |n
\rangle$ and $\langle m|\bar 
z |n \rangle$. Since $zP_n(z)$ can be
expressed through polynomials 
of the degree not grater than $n$, one may
represent $\langle m|z |n \rangle$ 
and $\langle m|\bar z |n \rangle$ in terms
of shifts  operators
$W=e^{\hbar\frac{\p}{\p t_0}}$
in the form of (\ref{L},\ref{L1}),  
where $r(t_0=\hbar n) = h_{n}/ h_{n+1} $.

Similar arguments allow one to identify the flows. Consider a 
variation
of some operator $\langle m|O |n \rangle$ under a variation of
$t_k$. We have $\hbar \p_{t_k}\langle m|O |n \rangle=\langle m[| H_k, O]|n
\rangle$, where  $H_k=A_{k} -
{A^\dag}_k$ and
$ \langle m| A_{k}|n \rangle= \langle m|\p_{t_k}|n \rangle.$ Obviously
$ H_{k}=-L^k(W)+$ negative powers of $W$. Choosing $O$ to be $\bar L$
(see (\ref{L1})) which consists on $W^{-1}$ and positive powers of $W$,
one concludes that $H_k$ does not consists of negative powers of $W$.
This brings us to  eq.(\ref{75}).

Finally, the operator
     $D=\langle m|\hbar \p_z|n \rangle$ is equal to
\beq
\label{worD}
D= \bar L  - \sum_{k\ge 1}  kt_{k} L^{k-1} , \quad
\eeq
The  Heisenberg relation $[D, L]=\hbar$ prompts the string equation
(\ref{string}).

The matrix model also offers an effective method to derive
eqs.(\ref{i}-\ref{Hirota}) (see e.g. \cite{2matrix}).

\bigskip
\noindent
{\it 10. $\tau$-function and spectral properties 
of the Dirichlet problem.}
This subject is under current study.

\bigskip\noindent
We thank M. Brodsky, V. Kazakov, S.P. Novikov 
and L.Takhtajan for valuable
comments and interest to this work.

The work of I.K.
is supported in part by European TMR contract ERBFMRXCT960012 and EC
Contract FMRX-CT96-0012.
The work of I.Kr. is supported in part by NSF grant DMS-98-02577.
P.W. would like to thanks P.Bleher and A.Its for the hospitality in
MSRI during the workshop on Random Matrices in spring 1999.
I.Kr. and A.Z. have been partially supported by
CRDF grant 6531.
P.W. and A.Z. have been partially supported by grants NSF DMR
9971332 and MRSEC NSF DMR 9808595. The work of A.Z.
was supported in part
by grant INTAS-99-0590 and RFBR grant 00-02-16477.
He also thanks for hospitality
the Erwin Schr\"odinger Institute in Vienna, where this
work was completed.

\end{document}